\documentclass[10pt,letterpaper]{article}
\usepackage[top=0.85in,footskip=0.75in]{geometry}
\usepackage{verbatim}

\usepackage{amsmath,amssymb}

\usepackage{numprint}

\usepackage{changepage}

\usepackage[utf8x]{inputenc}

\usepackage{textcomp,marvosym}

\usepackage{cite}

\usepackage{nameref,hyperref}

\usepackage[right]{lineno}

\usepackage{microtype}
\DisableLigatures[f]{encoding = *, family = * }

\usepackage[table]{xcolor}

\usepackage{array}
\usepackage{hhline}
\usepackage{float}

\usepackage[colorinlistoftodos]{todonotes}

\newcolumntype{+}{!{\vrule width 2pt}}

\newlength\savedwidth

\usepackage[aboveskip=1pt,labelfont=bf,labelsep=period,justification=raggedright,singlelinecheck=off]{caption}

\makeatletter
\renewcommand{\@biblabel}[1]{\quad#1.}
\makeatother

\usepackage{parskip}

\date{}

\usepackage{lastpage,fancyhdr,graphicx}
\usepackage{epstopdf}
\usepackage{footnote}
\makesavenoteenv{tabular}
\makesavenoteenv{table}

\usepackage{hyperref}
\usepackage{makecell}
\usepackage{cleveref}

\begin{document}
\vspace*{0.2in}

% Title must be 250 characters or less.
\begin{flushleft}
{\Large
\textbf\newline{The emerging sectoral diversity of startup ecosystems} % Please use "sentence case" for title and headings (capitalize only the first word in a title (or heading), the first word in a subtitle (or subheading), and any proper nouns).
}
\newline
% Insert author names, affiliations and corresponding author email (do not include titles, positions, or degrees).

Cl\'ement Gastaud\textsuperscript{1,2},
Théophile Carniel\textsuperscript{1,2},
Jean-Michel Dalle\textsuperscript{*,1,2,3}

\bigskip
\textbf{1} Agoranov, Paris, France

\textbf{2} Sorbonne Universit\'e, Paris, France

%\textbf{3} Laboratoire de Physique des solides, Universit\'e Paris Sud, Universit\'e Paris-Saclay, Orsay, France

\textbf{3} i3-CNRS, \'Ecole Polytechnique, France

\bigskip

% Insert additional author notes using the symbols described below. Insert symbol callouts after author names as necessary.
% 
% Remove or comment out the author notes below if they aren't used.
%
% Primary Equal Contribution Note
%\Yinyang These authors contributed equally to this work.

% Additional Equal Contribution Note
% Also use this double-dagger symbol for special authorship notes, such as senior authorship.
%\ddag These authors also contributed equally to this work.

% Current address notes
%\textcurrency Current Address: Dept/Program/Center, Institution Name, City, State, Country % change symbol to "\textcurrency a" if more than one current address note
% \textcurrency b Insert second current address 
% \textcurrency c Insert third current address

% Deceased author note
%\dag Deceased

% Group/Consortium Author Note
%\textpilcrow Membership list can be found in the Acknowledgments section.

% Use the asterisk to denote corresponding authorship and provide email address in note below.
* jean-michel.dalle@sorbonne-universite.fr %(J.-M. D.)

\end{flushleft}
% Please keep the abstract below 300 words
\section*{Abstract}
Thanks to the recent availability of comprehensive and detailed online databases of startup companies, it has become possible to more directly investigate startup ecosystems i.e. startup populations in specific regions. In this paper, we analyze the emergence of 20+ such ecosystems in Europe and the USA, with a specific focus on their sectoral diversity. Analyzing the sectoral landscapes of these ecosystems using a new visualization tool indeed highlights marked differences in terms of diversity, which we characterize using metrics derived from ecological sciences. Numerical simulations suggest that the emerging diversity of startup ecosystems can be explained using a simple preferential attachment model based on sectoral funding. 

% Please keep the Author Summary between 150 and 200 words
% Use first person. PLOS ONE authors please skip this step. 
% Author Summary not valid for PLOS ONE submissions.   
%\section*{Author summary}

% Use "Eq" instead of "Equation" for equation citations.
\section{Introduction}

Startup populations have recently come to be commonly referred to as \emph{"startup ecosystems"}, by analogy with ecological systems. 
This metaphor has emerged in economics and management sciences in the early 90's from different sources in order to study the creation, growth and death of organizations~\cite{Hannanh}, the competition between industrial actors~\cite{MooreHBR} or else the emergence of new technology niches~\cite{Schot}.
More recently, startup ecosystems have become central in local, national and international innovation policies~\cite{Gilbert, Minniti} as innovative startups were drawing increasing investments~\cite{NVCA} from venture capitalists and an increased attention from stakeholders notably because of their potential to create jobs~\cite{Birch,Thurik,Wong}. 
%From support programs~\cite{Hawaii} to tax incentives~\cite{Nevada}, startup ecosystems have indeed become strategic for local governments and national policymakers.
Indeed, following the leading example of San Francisco and the Silicon Valley, Austin, Boston, Los Angeles or New York in the US and Berlin, London or Paris in Europe~\cite{Slush} have thrived to become active entrepreneurial ecosystems and are competing against one another in order to attract startups.

In this context, data and models that would allow entrepreneurs, investors and policy makers to analyze, characterize and compare the emergence and dynamics of different startup ecosystems are however still mostly missing.
Even if professional websites such as~\cite{VS,TC,CBInsights} have started to gather relevant information, there is a global lack of understanding concerning the fundamental mechanisms driving the development of entrepreneurial ecosystems.
Startup landscapes only provide a representation of the startups in a specific sector, split in sub-sectors, such as the global fintech landscape edited in 2016 by Atherton Research~\cite{landscape_fintech} or startups associated with a specific technology and geographical zone, such as the 2017 France Is AI’s landscape~\cite{landscape_ai}, all the more so as existing landscapes are mostly instantaneous snapshots and lack completeness. 
Put differently, there does not exist, as far as we know, any quantitative model or tool that would significantly help actors make a more appropriate sense of the dynamics of startup ecosystems, a fact that is somewhat surprising since entrepreneurship has become a major topic in public policy decision making~\cite{meyer, Chandra}.

In this article, we argue that the recent availability of comprehensive public startup databases represents an opportunity for the formulation and validation of such theoretical models, related to the dynamics of startup ecosystems. Building upon an automated startup landscape generator that allows for the visualization of entrepreneurial ecosystems while incorporating relevant metadata (e.g. textual descriptions, sector of activity and funds raised) \cite{gastaud}, we first suggest to extend the ecological analogy and characterize ecosystems using diversity metrics. We then try to relate the observed differences among ecosystems to macro-economic indicators before presenting and calibrating a numerical simulation that explains the diversification of startup ecosystems with a preferential attachment model based on the funding received within each sector.

\section{Materials and methods}
\subsection{Dataset}

\begin{figure}[H]
  \begin{center}
  \includegraphics[width=14cm]{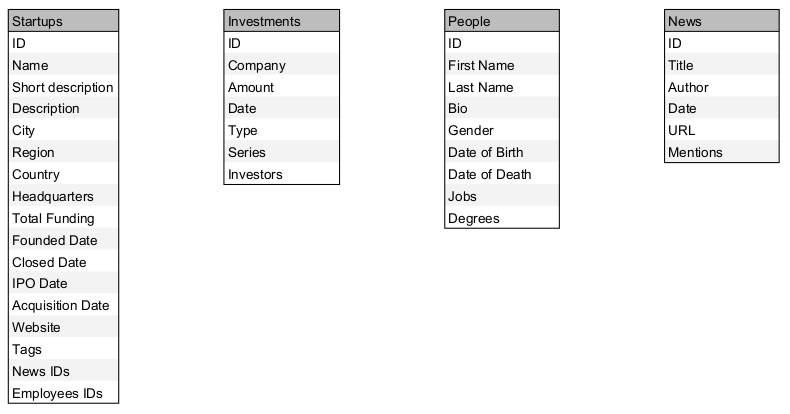}
  \caption{{\bf Data recovered from Crunchbase.}}
  \label{fig:CrunchbaseData}
  \end{center}
\end{figure}

% For figure citations, please use "Fig" instead of "Figure".
We exploit a dataset of startups from Crunchbase~\cite{CB}, a mainstream source of data for academic research with respect notably to US startups~\cite{dalle2017}. 
For European ecosystems, this dataset was supplemented by Dealroom~\cite{DR}, which increased the number of considered companies by 9.4\%.
For each startup, we retrieved its date of creation, location, sectoral tags (describing its economic sector, technology and/or market), textual description and, most notably, all the information with respect to the funds that the startup has raised, including the date at which they were raised, the amount of funding, the nature of the funding round and the identity of the investors as well as all the articles mentioning this company available on Crunchbase (Figure~\ref{fig:CrunchbaseData}).
In addition, we retrieved all information available about people on Crunchbase, giving us in particular proxies with regard to the experience of startup founders.
By nature of the funding round, we mean the different stages of Venture Capital funding that startup companies go through.
In this respect, the first round of funding is generally called \emph{Seed} and corresponds to money used to validate that the product of the startup and the market are in phase.
Other rounds are labeled by letters: \emph{A, B, C}, etc.
\emph{A} rounds are designed to ensure the scalability of the company while later rounds (\emph{B} and latter ones) tend to accompany the growth of the company in national and international markets~\cite{VCrounds}.
We limited our sample to companies created after January 1st, 1998 and to companies that mentioned at least one round of funding. 
Overall, our dataset consists in $\numprint{618366}$ companies, $\numprint{221299}$ investment rounds, $\numprint{783787}$ people and $\numprint{6363831}$ news articles.

For each startup, we further computed two additional metrics: first, the total amount of funds it has raised and second, the speed at which these funds were raised (as a proxy for its pace of growth) which we denote as its momentum and define at time $t$, in dollars per month, as:

\begin{align}
    \label{eq:vitality}
	m(t) = \sum_{i \in \{\text{investments}\}} \frac{\text{MoneyRaised}_{i}(t) \times \exp({-\text{DaysSinceInvestment}_{i}(t)/{365 \text{~days}}})}{12\text{~months}}
\end{align}

\subsubsection{Construction of the sectoral tree}

In order to visualize ecosystems, we organized startups according to their main economic sectors. 
We used Crunchbase's basic tag structure as a starting point to create a startup sectoral tree. 
This tag structure is organized in two levels: first, the industry (\textit{Health Care, Software}...) and then a more specific level (\textit{Health Insurance, Image Recognition, Construction}...). 
The resulting ontology was cleaned up by removing tags that were specially broad and not distinctive (\textit{Software, Infrastructure}, etc.), unrelated to economic sectors (\textit{B2B, Freemium}, etc.) or very rare (e.g. \textit{Ports and Harbors} that was only associated with 10 startups worldwide in our dataset). 
Furthermore, tags that were semantically very close (e.g. \textit{Shipping} and \textit{Delivery}, \textit{Video Games} and \textit{Gaming}...) were merged.
Then, whenever two tags had an inclusion relation not taken into account in the initial ontology (e.g. \textit{Insurance} and \textit{Health Insurance}), this relation was used to create a new sub-level in the tree, as in:

\begin{center}
    \textit{Financial Services → Insurance → Health Insurance}
\end{center}

\begin{figure}[!h]
    \centering
    \includegraphics[scale = .6]{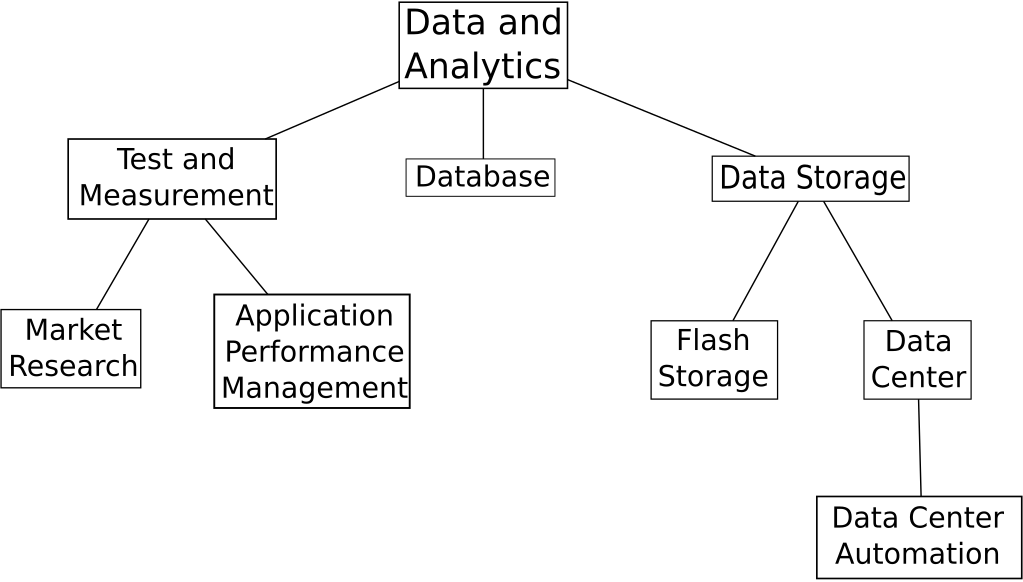}
    \caption{Part of the \textit{Data \& Analytics} industry subtree of the sectoral tree.}
    \label{fig:ontology_extract}
\end{figure}

It should be noted that in a few cases, visualization and classification prompted a manual edit of sectoral tags that were found to be either imprecise (e.g. startups with only a very general tag such as \textit{Software}) or too numerous (e.g. a startup tagged with all the industries that could possibly make use of its technology) or simply factually erroneous.
Following this procedure, the final sectoral tree was composed of 478 sectoral tags, down to 4 levels and composed of 28 industries i.e. independent sectors directly connected to the root of the tree\footnote{A full description of the sectoral tree is available upon request to the authors.}. Part of the \textit{Data and Analytics} branch is shown as an example in fig.~\ref{fig:ontology_extract}.

\subsubsection{Populating the sectoral tree with startups}

This sectoral tree is used to populate a startup tree, considering startups as end leaves.
%Thanks to the tags of each startup, each one of them can be precisely positioned at the right place in the tree.
Since most startups have several sectoral tags, we implemented a heuristic procedure to prune the tree i.e. to keep the most relevant tag for each startup.
%There is no obvious preferred selection process, thus we chose to follow an optimization strategy.
Following a strategy similar to~\cite{batista2015}, we can determine a startup's main industry (or tag) by classifying its description. For each tag, we compute the probability that a startup is best described by it. If this startup has no tag, we choose the most probable tag overall. If it has several tags and some are included in others, we first remove the shallowest ones in the sectoral tree as it corresponds, by definition, to the least precise sectoral assignment. Then, we choose the most probable tag from the remaining ones.

In the end, the simplest (least ambiguous) possible sectoral tree is obtained with startups associated as end leaves.

\subsubsection{An interactive visualization tool for startups ecosystems}

In order to visualize the ecosystems, we made use of the TreeMaps FoamTree package~\cite{FoamTree}, which allows to display hierarchical data as nested polygons tiling the plane, each cell having a surface proportional to a specific dimension of the data, as is general in tessellations and treemap representations~\cite{bib1}.

Examples of such visualizations are presented in appendix. Each cell of the map corresponds to a startup, its surface representing the amount of funding received by the startup and its color the momentum as defined in eq.~\ref{eq:vitality}. The visualization typically confirms widely acknowledged characteristics of these ecosystems: for instance, London appears specialized in FinTech (22.6\% of the investments) while Paris appears particularly strong with respect to Health Care.
Furthermore, each ecosystem can be easily visualized through an interactive interface~\footnote{Available at: \url{http://atlas.agoranov.com}}, while several filters can be applied to the map using all the data available on startups: tags, location, investors, etc.
Thanks to the timestamps on each event, an ecosystem can also be visualized at any given date in order to study ecosystem and investment dynamics.

\begin{comment}
would like to visualize two things: where they are now and where they could be in the future.

With the data available, an easy way to apprehend the \textit{"position"} of a startup is how much money they have raised. 
This metric gives an idea of how advanced and trusted the project is by investors while being easily accessible and understandable.
Concerning the future of a startup, there is not such a clear metric. 
An idea of the \textit{"speed"} of a startup could be used to determine a trajectory. 
Since the position is the funding, speed can be reasonably compared to the \emph{cash burn} of the company, \textit{i.e.} how much money the start-up spend per unit of time. 
Indeed, it is directly correlated to the number of employees and the material investments.

However, the cash burn of a company is not directly available. 
Therefore, a proxy metrics is defined: the \emph{“vitality”} $v$ of a startup.
The vitality depends on the funding amount and the time since the funding rounds in dollars per month. 
The more a startup has raised recently, the greater the vitality will be. 
\end{comment}

\subsection{Introducing diversity metrics for startups ecosystems}

To better characterize startup ecosystems, we introduce diversity metrics similar to what is traditional for ecological ecosystems.
In ecological ecosystems, diversity is on average positively correlated with stability~\cite{McCann}: if a change in a diverse environment (for example a disease, or the arrival of a predator) targets some species, the impact on the whole ecosystem will be reduced because of functional redundancy.
In economics, the relationship between diversity and unemployment stability has been widely studied~\cite{Gilchrist,Dissart} and it has notably been proposed that, as for ecology, diversity was positively correlated with stability through resilience of the economy to rapid changes~\cite{Kurre,Kort,Dissart}, although empirical analysis using regional data does not always confirm this hypothesis~\cite{Dissart}.
Similarly, a disruption in some sectors might more or less affect an entire startup ecosystem depending on its diversity across industries and sectors.

%with a technology becoming obsolete or a new innovative competitor emerging as it has been the case with Uber and cabs or AirBnB and hospitality for instance. 
%If a city bases most of its economy on a specific industry and it gets disrupted, it will be harder to recover economically from the shock.
%Based on this insight, diversity of several ecosystems has been studied and explanations of the disparities between them have been proposed.

At least three major diversity indices have been defined and used in ecology: the Simpson index, the Shannon-Weiner index and the Hill index~\cite{Pielou}. Both the Shannon-Weiner and Simpson indices and their corresponding diversity can be derived from the Hill numbers of order $q = 1$ and $q = 2$ respectively~\cite{Hill1973}. In the context of this study, we implemented the Shannon-Weiner index (measures the diversity of the ecosystems within the previously defined sectoral ontology) and the Herfindahl-Simpson index (measures the concentration of investments between startups regardless of their sectors and industries). 

\subsubsection{The Simpson \& Herfindahl indices}
\label{subsec:simpson_index}
In ecology, studies usually present the Simpson index~\cite{simpson} defined as:

\begin{align}
    \label{lambda}
	\lambda = \sum_{i=1}^{N} p_i^2
\end{align}

where traditionally in ecology, $N$ is the total number of species and $p_{i}$ the relative abundance of the species $i$.
In the present case $N$ would be the total number of tags and $p_{i}$ the ratio between funding invested in sector $i$ and the total funding of the ecosystem. 
The Simpson index measures the probability that two individuals randomly chosen from a population belong to the same species.
The extreme cases $\lambda = \frac{1}{N}$ and $\lambda = 1$ correspond respectively to a maximal and a minimal diversity.

In order to have a more straight forward interpretation, the inverse Simpson index $1/\lambda$ is often used. This corresponds to the \textit{effective number of species} or \textit{true diversity} $D$ as defined in~\cite{Jost2006, Jost2007}. To give a quick intuition of the concept, this number converts the computed diversity index of the studied ecosystem into a corresponding ecosystem where all species are equally abundant; the resulting number of different species corresponds to the effective number of species (an unbalanced ecosystem with $M$ species each with different values of $p_{i}$ would be converted into an ecosystem with $N \leq M$ species each with $p_{i} = \frac{1}{N}$ for $i \in [1, N]$).

This index is also used in economics where it is called the Herfindahl index~\cite{herfindahl} and is usually used to study the importance of a company on a given market.
It is defined as follows :
\begin{align}
    \label{eq:H}
	H = \sum_{i=1}^{N} s_i^2
\end{align}
where $s_i$ is the market share of a company $i$.

However, this index does not take the sectoral tree structure into account and focuses solely on the repartition of funds between actors. 

\subsubsection{The Shannon-Weiner index}
\label{subsec:shannon_index}
The Shannon-Weiner index, or Shannon entropy, originating from information theory~\cite{Shannon} and statistical physics~\cite{Gibbs}, is defined as follows:

\begin{align}
    \label{eq:entropy}
	S = - \sum_{i=1}^{N} p_{i} \log_{b} p_{i}
\end{align}

In the ecology literature, base $e$ for the logarithm is usually used~\cite{Jost2006}; this convention will be used in the following. Shannon entropy quantifies the uncertainty associated with the prediction of an element of the considered dataset. In the context of ecology, it quantifies the uncertainty in predicting to which species an individual taken at random from the dataset belongs. However, in this form, the additional information given by the tree structure of the data is still not taken into account. Its hierarchical structure needs to be taken into consideration in the analysis. A more apt measure of the entropy $S_{\tau}$ of a tree $\tau$ is thus :

\begin{align}
    \label{eq:entropy2}
	S_{\tau} = - \sum_{\beta=1}^{N_{\tau}} p_{\beta|\tau} (\ln p_{\beta|\tau} - S_{\beta})
\end{align}

with $N_{\tau}$ the number of branches originating from $\tau$, $S_{\beta}$ the entropy of the subtree $\beta$ and $p_{\beta|\tau}$ being either the ratio of funding invested in $\beta$ compared to total funding invested in $\tau$ or the ratio of the number of startups in $\beta$ compared to the total number of startups in $\tau$ (\textit{i.e.} the probability of $\beta$ knowing $\tau$). We refer to the entropy computed using the ratios of funding as \textit{Shannon funding} and the entropy computed using the ratios of number of startups as \textit{Shannon startups}. 

Naturally, this measure is dependent on the structure of the ontology defined previously. This issue is well-known in ecology and emerges from the definition of a species that one chooses to use~\cite{Agapow}.

Following~\cite{Jost2006, Jost2007}, the effective number of species $D$ can be derived from the Shannon-Weiner entropy index : 

\begin{align}
    \label{eq:shannon_diversity}
    D = \exp{(S_{\tau})}
\end{align}

The Shannon-Weiner index value for a tree with all categories having equal population (478 categories in total) is about $S_{\tau} = 6.1696$. The corresponding effective number of species is then $D = \exp(S_{\tau}) = 478$ which is coherent with our definition and understanding of this metric.

Hill numbers of order $q = 1$ (Shannon-Weiner diversity) are to be favored when calculating diversities without any prior information about the ecosystem (from~\cite{Jost2007}, "orders higher than 1 are disproportionately sensitive to the most common species, while orders lower than 1 are disproportionately sensitive to the rare species."). Upon applying Hill diversity indices of order-$1$ and -$2$ to our dataset, we indeed find that the order-$1$ index allows us to gain insight into ecosystem dynamics whereas the order-$2$ index does not discriminate well between ecosystems. We will use the order-$1$ index as our diversity measure in the following. 

\begin{comment}
In our case, it is even more complex since several parameters are important when studying startup diversity: industry, technology and targeted market. 
Some industries can use different technologies and some technologies can be applied to several industries, but both can be disrupted. 
In particular, information and communication technologies are omnipresent, while biotechnology is often restricted to one application. 
Thus, this study is focused only on industry as it is often complicated to retrieve information concerning the technology used by the startups.
\end{comment}

\subsection{Simulating ecosystem growth}
\label{subsec:sim_procedure}
To try and understand some of the mechanisms behind the growth and diversification of an entrepreneurial ecosystem, we simulated the development of a startup ecosystem as described by our ontology (number of startups and amount of funding in structured categories). The incremental populating of the ecosystem was done following a simple preferential attachment model on the current state of the tree. Two main variants of the model were used :

\begin{itemize}
    \item In the first one, the new startup is placed in category $i$ with $i \in [1, N]$ with probability
    
    \begin{align}
        \label{eq:PA_startups}
        p_i = \frac{(n_{i} + 1)^{\alpha_c}}{\sum_{j = 1}^{N}(n_{j} + 1)^{\alpha_c}}
    \end{align}
    
    where $n_i$ is the number of startups in each category and $\alpha_c$ a free parameter of the model.
    \item In the second one, the new startup is created with a funding amount drawn from a powerlaw distribution with exponent $\beta = -2$ and support $[x_{min} = 10^5, x_{max} = 10^9]$. This new startup is then placed in category $i$ with $i \in [1, N]$ with probability
    
    \begin{align}
        \label{eq:PA_funding}
        p_i = \frac{(f_{i} + x_{min})^{\alpha_c}}{\sum_{j = 1}^{N}(f_{j} + x_{min})^{\alpha_c}}
    \end{align}
    
    where $f_i$ is the total funding of the startups in category $i$ and $\alpha_c$ a free parameter of the model.
\end{itemize}

% Results and Discussion can be combined.
\section{Results and discussion}

\begin{figure}[!ht]
    \centering
    \includegraphics[width=15cm]{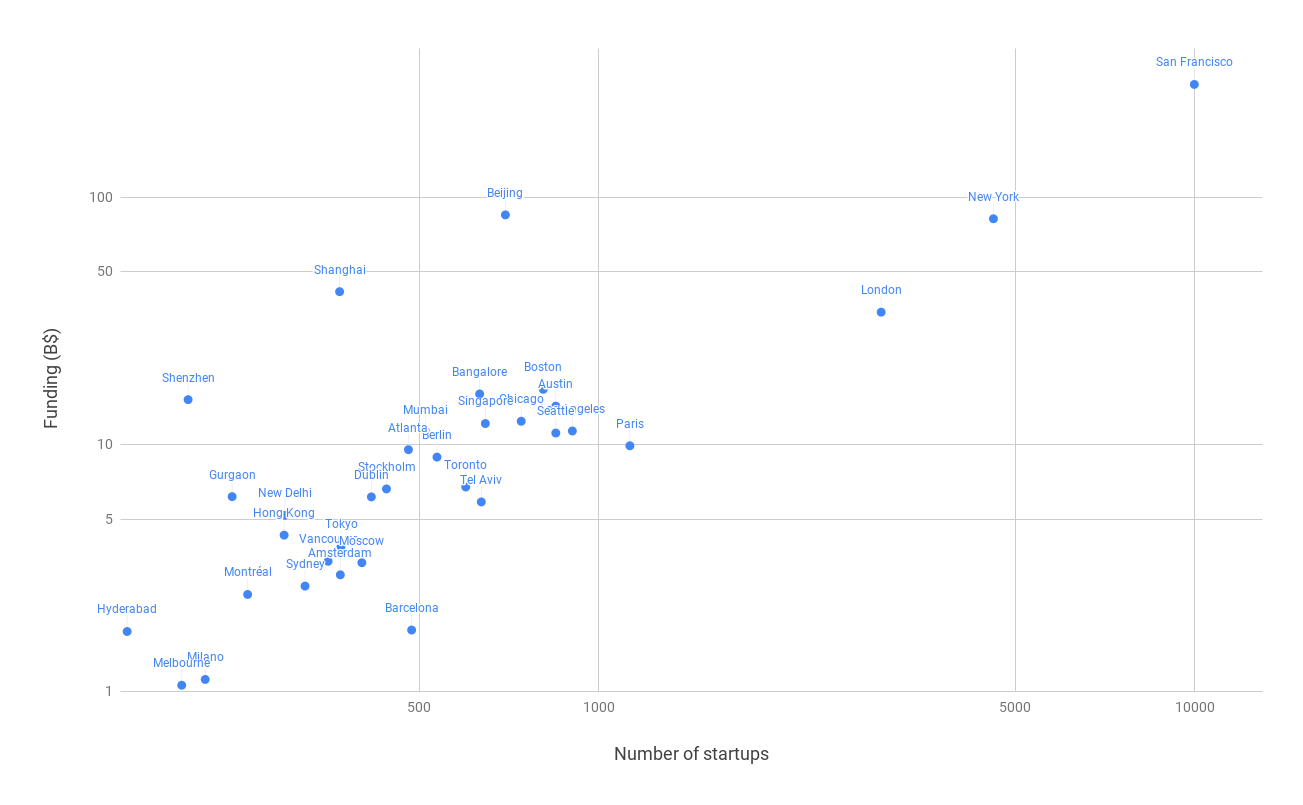}
    \caption{Ecosystem sizes in terms of number of startups and total funding.}
    \label{fig:ecosystems}
\end{figure}

\begin{figure}[!ht]
  \begin{center}
  \includegraphics[scale=0.69]{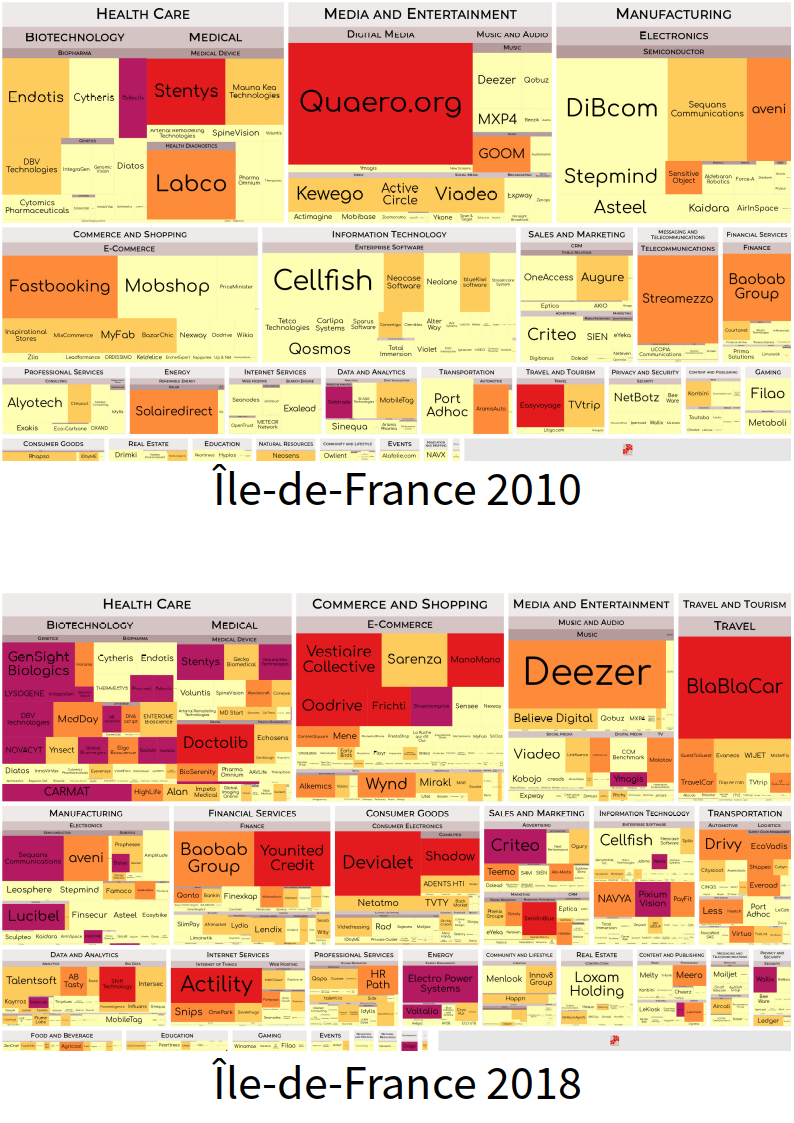}
  \end{center}
  \caption{\bf Maps of Île-de-France in 2010 and 2018 Jan 1\textsuperscript{st}.}
  \label{MapsParis}
\end{figure}

\begin{figure}[!ht]
  \begin{center}
  \includegraphics[scale=0.69]{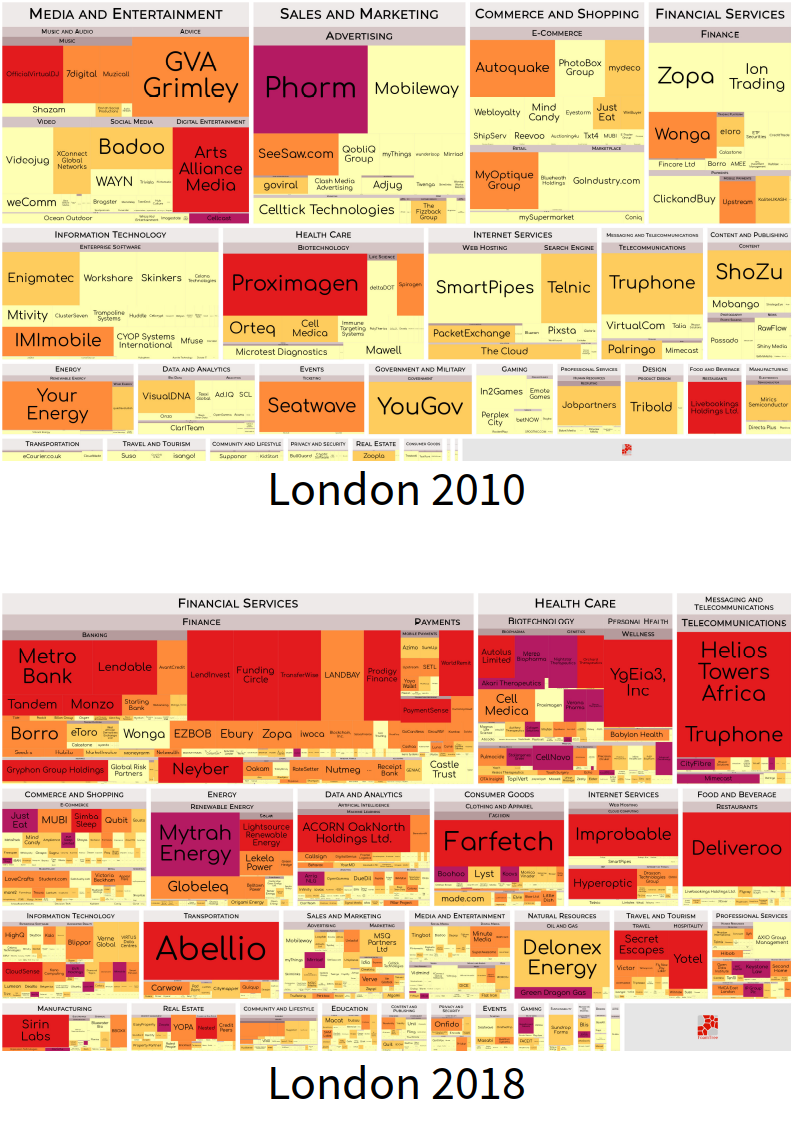}
  \end{center}
  \caption{\bf Maps of London in 2010 and 2018 Jan 1\textsuperscript{st}.}
  \label{MapsLondon}
\end{figure}

\begin{figure}[!ht]
  \begin{center}
  \includegraphics[scale=0.69]{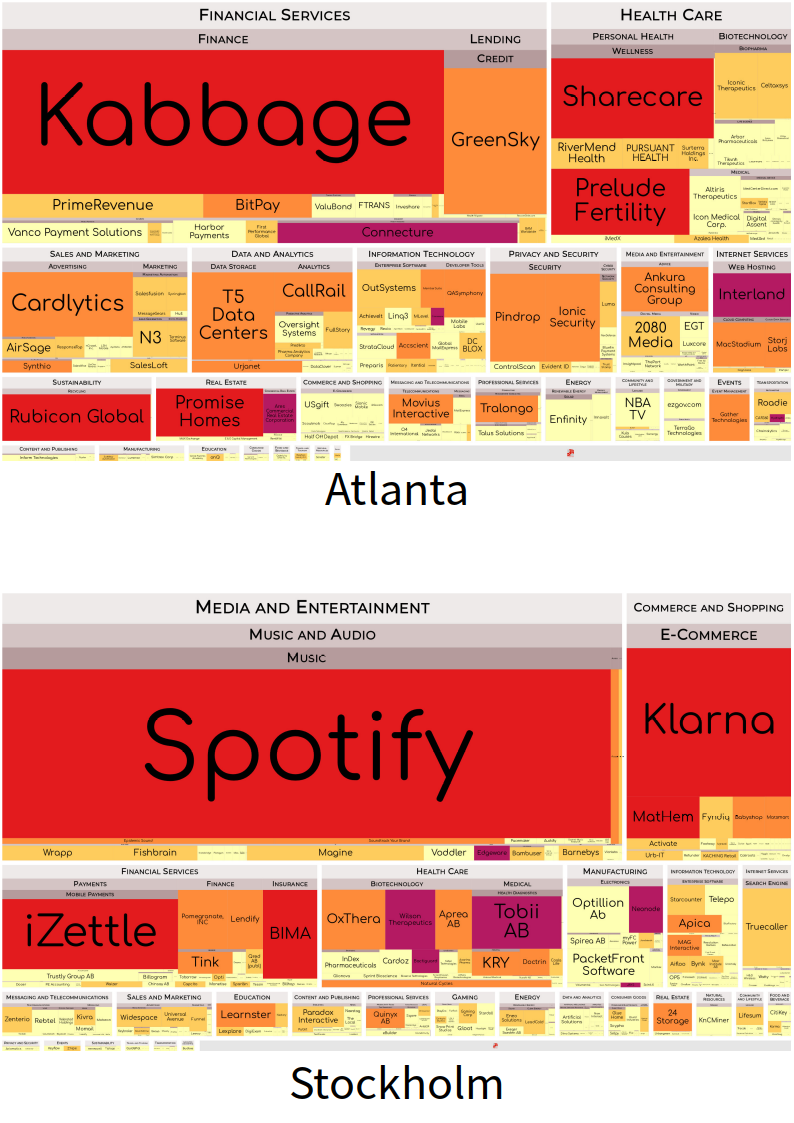}
  \end{center}
  \caption{\bf Maps of Atlanta and Stockholm on 2018 Jan 1\textsuperscript{st}.}
  \label{MapsATLSTO}
\end{figure}

\begin{figure}[!ht]
  \begin{center}
  \includegraphics[scale=0.69]{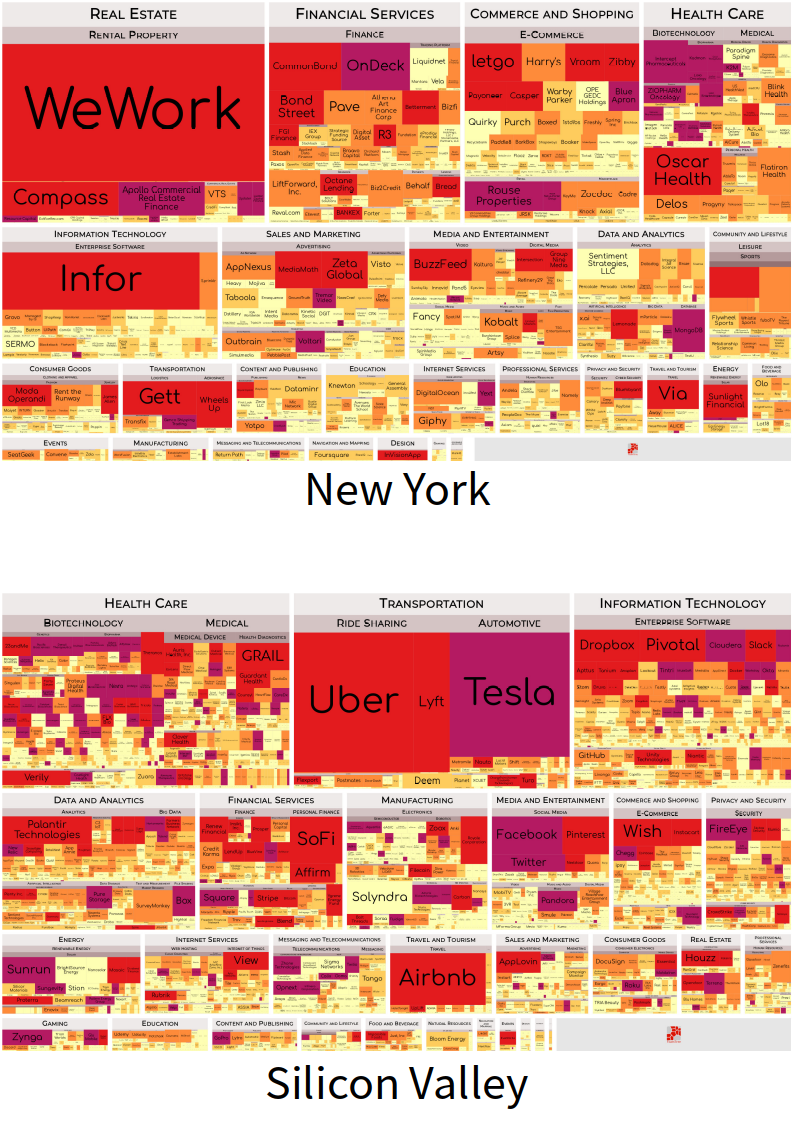}
  \end{center}
  \caption{\bf Maps of New York and the Silicon Valley on 2018 Jan 1\textsuperscript{st}.}
  \label{MapsNYSV}
\end{figure}

We applied our methodology to compare thirty-four ecosystems in Europe, North-America, Asia and Australia, chosen based on their prominence~\cite{Slush}.
Figure~\ref{fig:ecosystems} sums up the sizes in terms of number of startup companies and total funding of these ecosystems as of January the 1\textsuperscript{st}, 2018 and ~\cref{MapsParis,MapsLondon,MapsATLSTO,MapsNYSV} shows the mapping of some ecosystems using our visualization method. 
The size of the startup cells is proportional to the amount of funds they raised and the color encodes its momentum, as defined in Eq.~\ref{eq:vitality}. Purple means that the company went public, and the shades from red to beige represent in sequence the top 1\%, 5\% and 15\% startups with the highest momentum. The industries are ordered from top left to bottom right following their total funding.

As expected, the funding increases with the number of startups. However, the ecosystems visually exhibit significant disparities in terms of funding allocation. 
For instance, while Paris hosts twice as many startups as Atlanta, the total cumulative investments in both cities are comparable (since January 1st, 2000 \$9.5B in Atlanta vs \$9.9B in Paris). 
Mapping the ecosystems might shed some light on this observation. Some, like Atlanta, Berlin or Stockholm (fig~.\ref{MapsATLSTO}) appear characterized by a relatively weak diversity, related to the presence of a few champions -- \emph{unicorns} -- such as Kabbage in Atlanta, Delivery Hero in Berlin or Spotify, Klarna and iZettle in Stockholm that have raised billions of dollars. On the other hand, Paris, New York or the Silicon Valley (\cref{MapsParis,MapsNYSV}) appear much more diverse, in terms of funding as well as industry.

Being able to visualize the evolution of ecosystems also captures dynamic trends. For instance, the slow fall of \textit{Manufacturing} in Ile-de-France is explicit in these representations, falling from 12.7\% of the total investments in 2010 to 7.0\% in 2018 (fig.~\ref{MapsParis}). In London on the other hand, \textit{Financial Services} investments have skyrocketed over the same period from 10.3\% of investments to 22.6\% (fig.~\ref{MapsLondon}). It is now the biggest funding recipient in the British metropolis.

\subsection{Measured diversity}
\label{subsec:measured_diversity}
\begin{figure}[!ht]
    \centering
    \includegraphics[scale = .4]{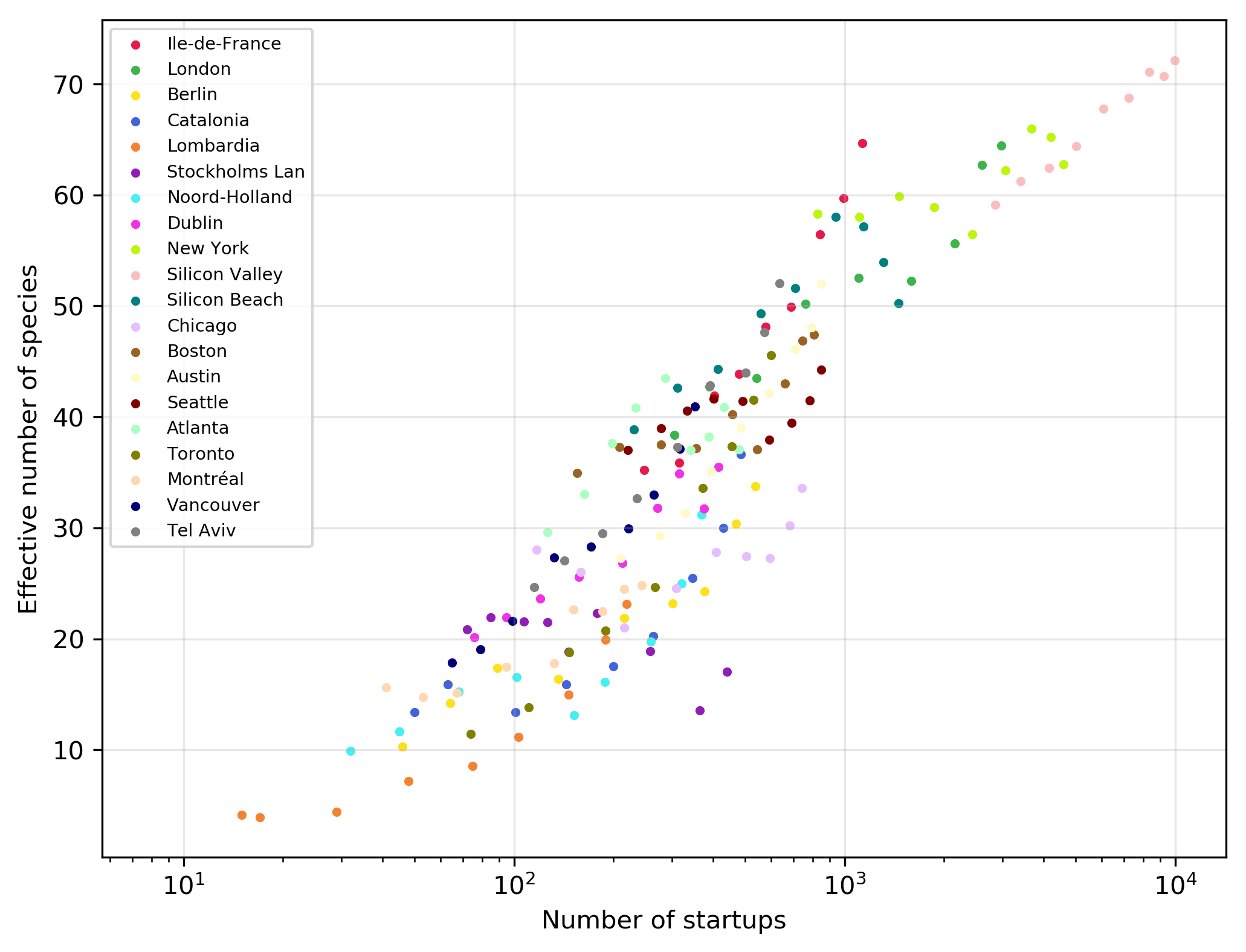}
    \includegraphics[scale = .4]{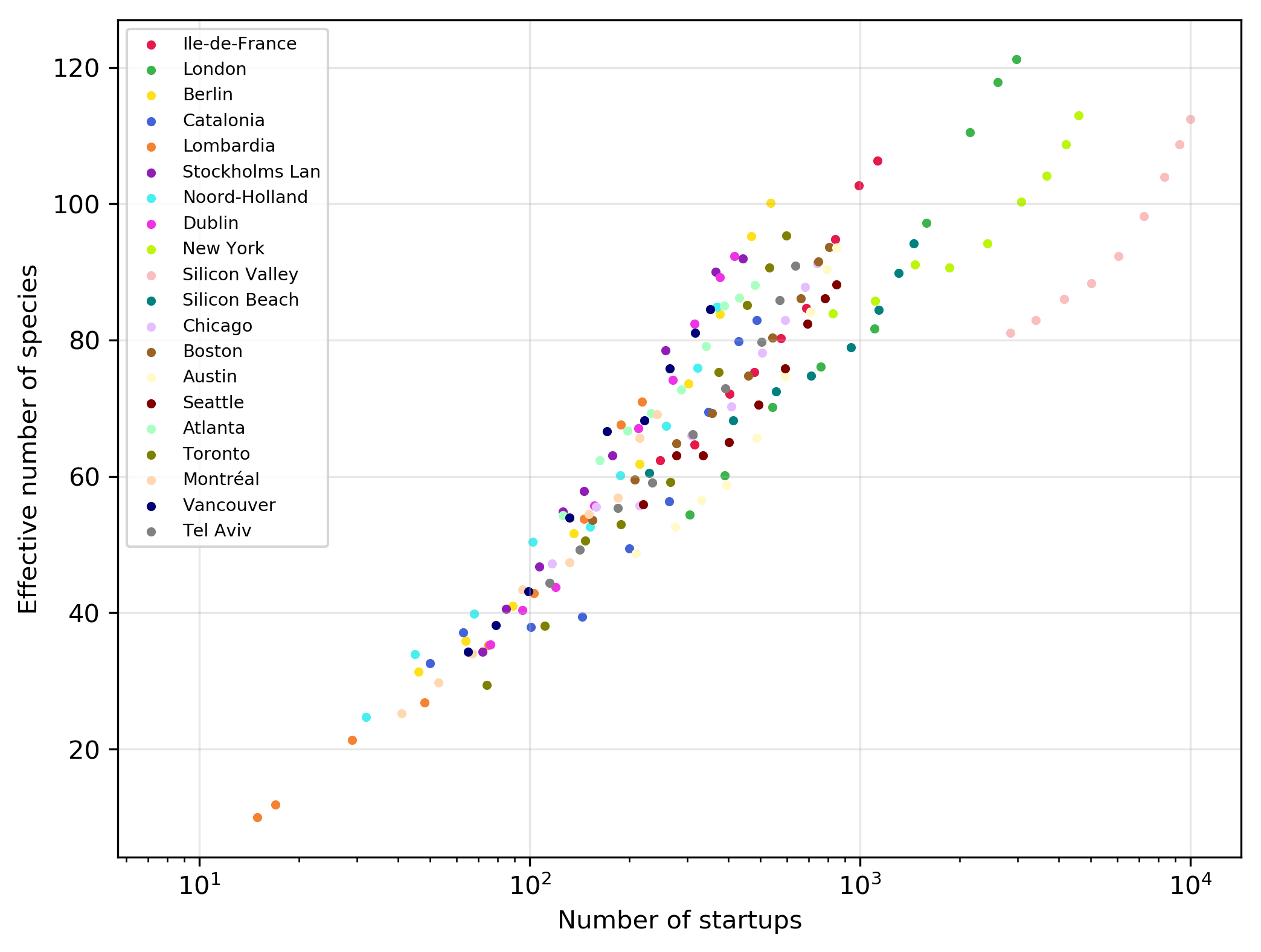}
    \caption{Temporal evolution of the effective number of species as a function of the number of startups for various ecosystems between 01/01/2010 and 01/01/2018. Values of the effective number of species were computed using the entropy calculated on the funding in each category (left) and the number of startups in each category (right).}
    \label{fig:data_ecosystems}
\end{figure}

Fig.~\ref{fig:data_ecosystems} shows the evolution of selected ecosystems in terms of number of startups and effective number of species between January the 1\textsuperscript{st}, 2010 and January the 1\textsuperscript{st}, 2018. Each of them is represented by a pair of values $\{ N(t), D(t) \}$ per year with $N(t)$ and $D(t)$ respectively the number of startups in the ecosystem and the effective number of species in the ecosystem at time $t$. Diversity is computed using the funding per category (Shannon funding) for the left plot and the number of startups per category (Shannon startups) for the right plot. Diversity is higher for Shannon startups compared to Shannon funding and individual trajectories tend to be more distinct for high numbers of startups (see for instance Silicon Valley or New York), suggesting that ecosystem-specific dynamics could be at play when the ecosystem becomes sufficiently large.

\begin{figure}[!ht]
    \centering
    \includegraphics[scale = .4]{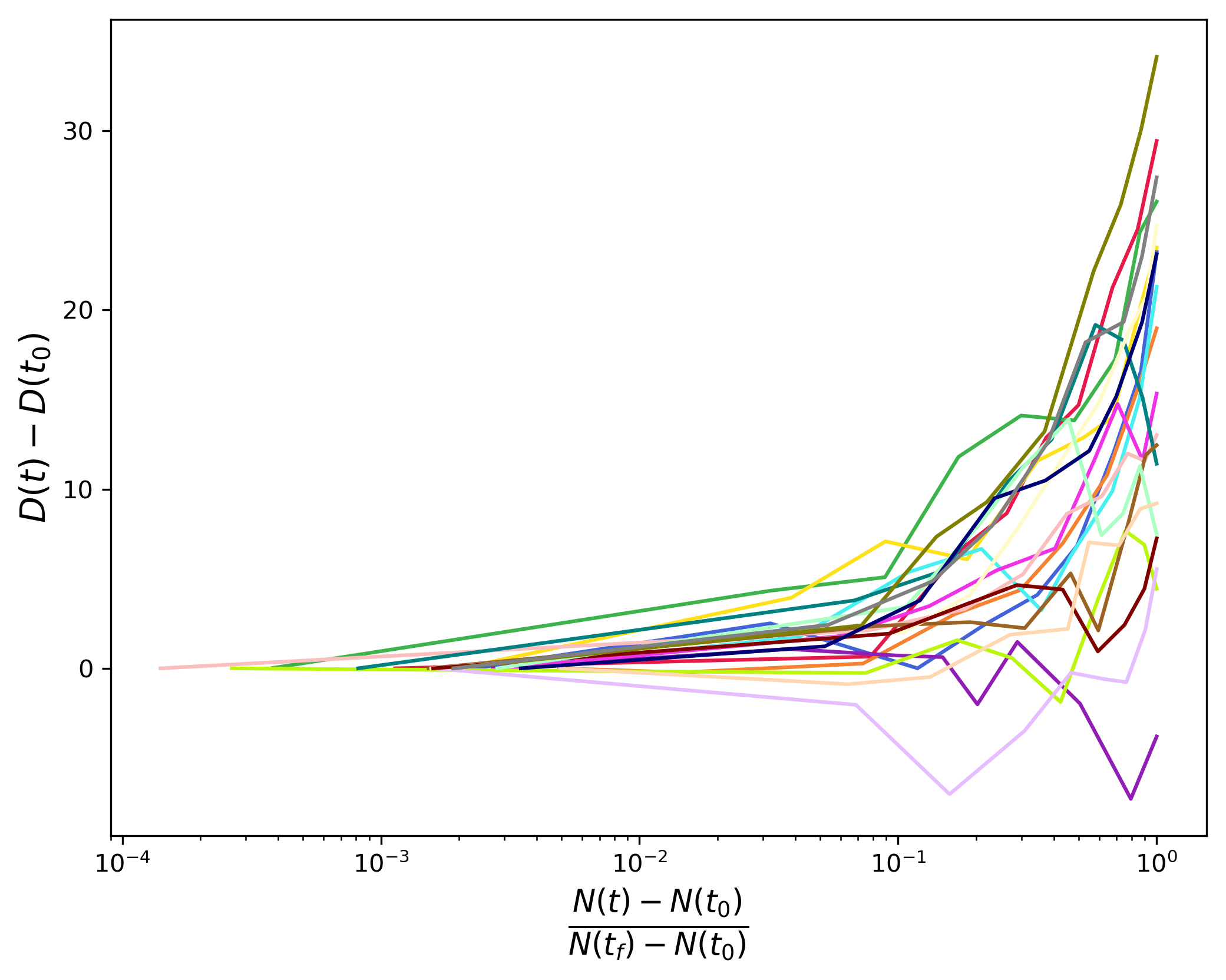}
    \includegraphics[scale = .4]{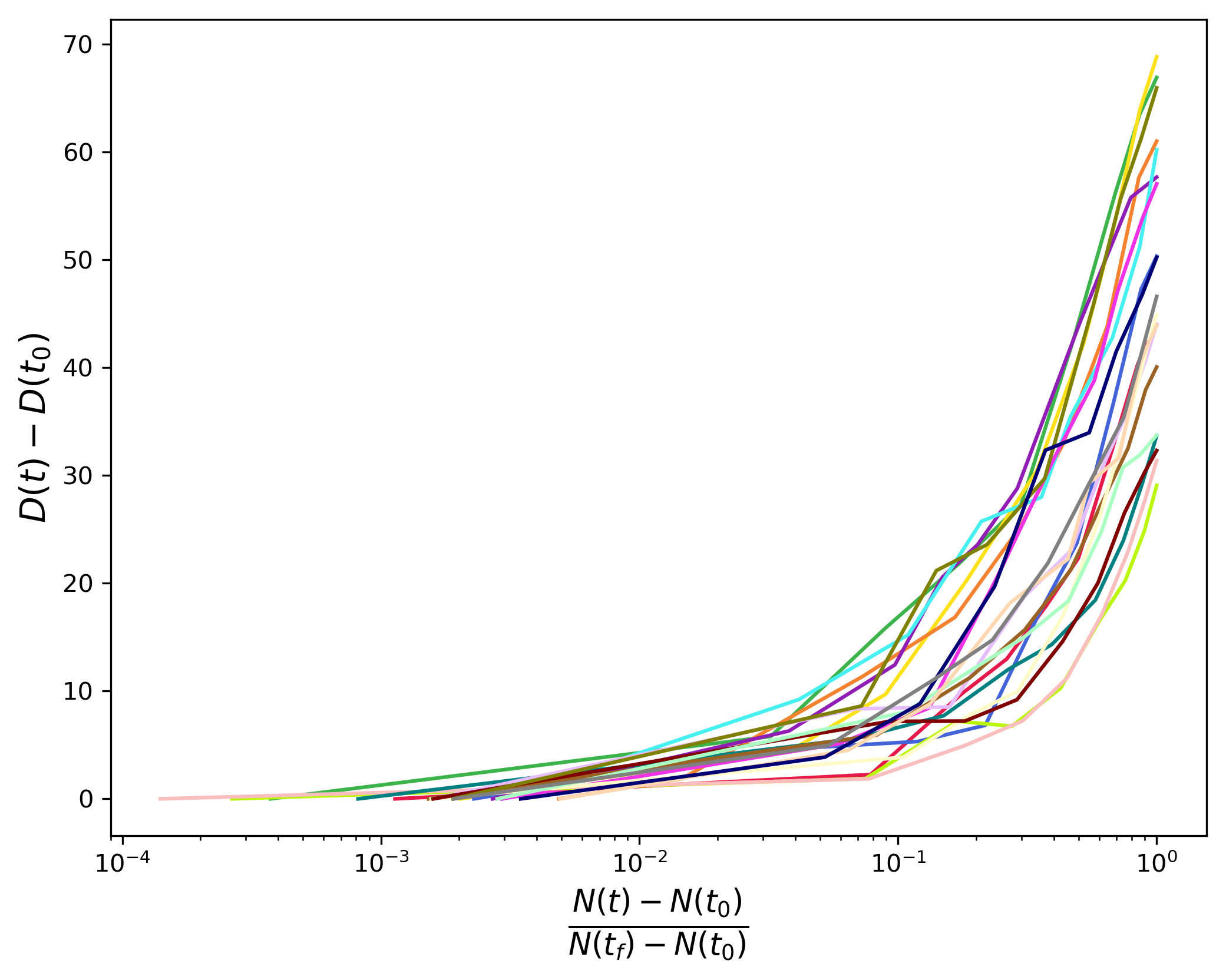}
    \caption{Standardized temporal ecosystem trajectories. $\hat{D}(t)$ is the effective number of species and $\hat{N}(t)$ is the number of startups at time $t \in [0, t_{f}]$. Values of the effective number of species were computed using Shannon funding (left) and Shannon startups (right).}
    \label{fig:std_data_ecosystems}
\end{figure}

Since ecosystems differ widely in terms of number of startups, it is useful to scale diversity trajectories so that the number of startups at the start and end of the measuring period are comparable. Fig.~\ref{fig:std_data_ecosystems} presents the standardized ecosystem dynamics \textit{i.e.} ecosystem are characterized by value pairs $\{ \hat{N}(t), \hat{D}(t) \}$ :

\begin{align}
    \label{eq:standardization}
    \hat{N}(t) = \frac{N(t) - N(t_0)}{N(t_f) - N(t_0)}\\
    \hat{D}(t) = D(t) - D(t_0)
\end{align}
with $t_0$ the index of the first data point (year 2010) and $t_f$ the index of the last data point (year 2018). 

Using these standardized metrics, all ecosystems have similarly-shaped trajectories using Shannon startups (right plot) whereas trajectories computed using Shannon funding (left plot) seem to be more variable, probably due to the large discrepancies in individual funding amounts which can easily unbalance an ecosystem especially in early stages of development. The diversification in terms of number of startups per sector (Shannon startups) thus seems to be a more fundamental characteristic shared between all our studied ecosystems when compared to the diversification in terms of funding per sector. We will therefore use Shannon startups to compute entropy and diversity during the numerical simulations.

\subsection{Correlations to macro-economic indicators}
\label{subsec:macro_indicators}
In order to move beyond visual intuitions from the landscapes, we made use of the diversity metrics defined in the previous section.
We fitted an OLS model to find correlations between the effective number of species $D$ and macro-economic indicators retrieved from the OECD Regional Statistics~\cite{oecd} including :
\begin{itemize}
    \item Wealth (GDP and GDP per capita),
    \item Economic vitality (Employment, GDP growth (base 2007)),
    \item Research intensity (\% labor force with tertiary education, number of researchers per 1000p, number of patents, R\&D expenses in M\$ and \% of GDP).
\end{itemize}

All the values are standardized by removing the mean and scaled to unit variance.
As the logarithm of the number of startups explain 80\% of the variance of $D$ ($R^{2} = 0.803$), we fit the indicators against $D(t) / log(N(t)$.
By fitting this value against the previously defined indicators while still controlling for the logarithm of the number of startups, we find a correlation with the GDP per capita (p-value of $1 \cdot 10^{-3}$).
Diversity thus only seems related to the economic development and, surprisingly, not at all to the research intensity of the metropolis. However, this observation can simply be a consequence of a higher maturity of startup ecosystems in developed countries, since they have existed for a longer time.

\subsection{Simulation results}
\label{subsec:sim_results}

\begin{figure}[!ht]
    \centering
    \includegraphics[scale = .4]{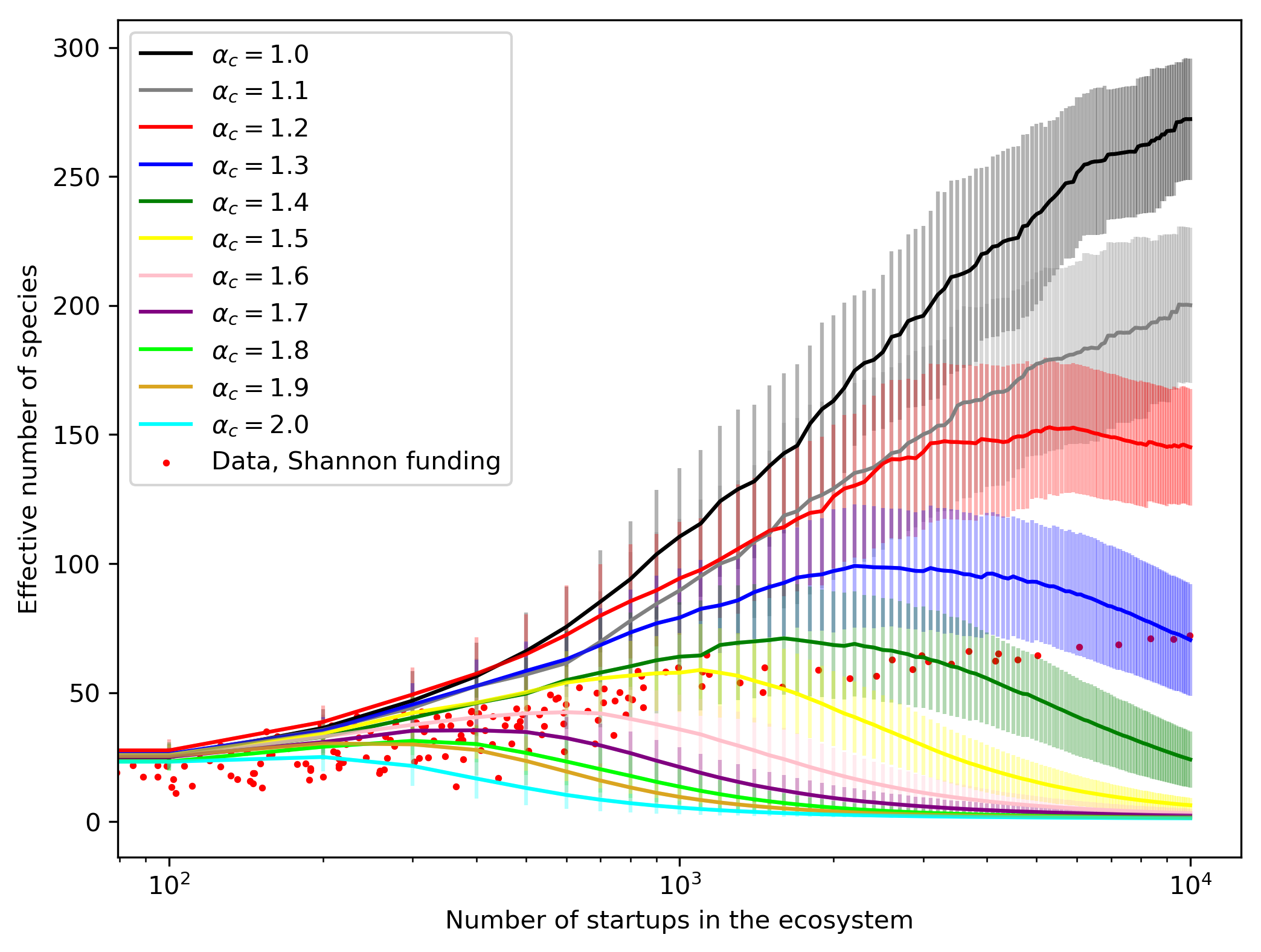}
    \includegraphics[scale = .4]{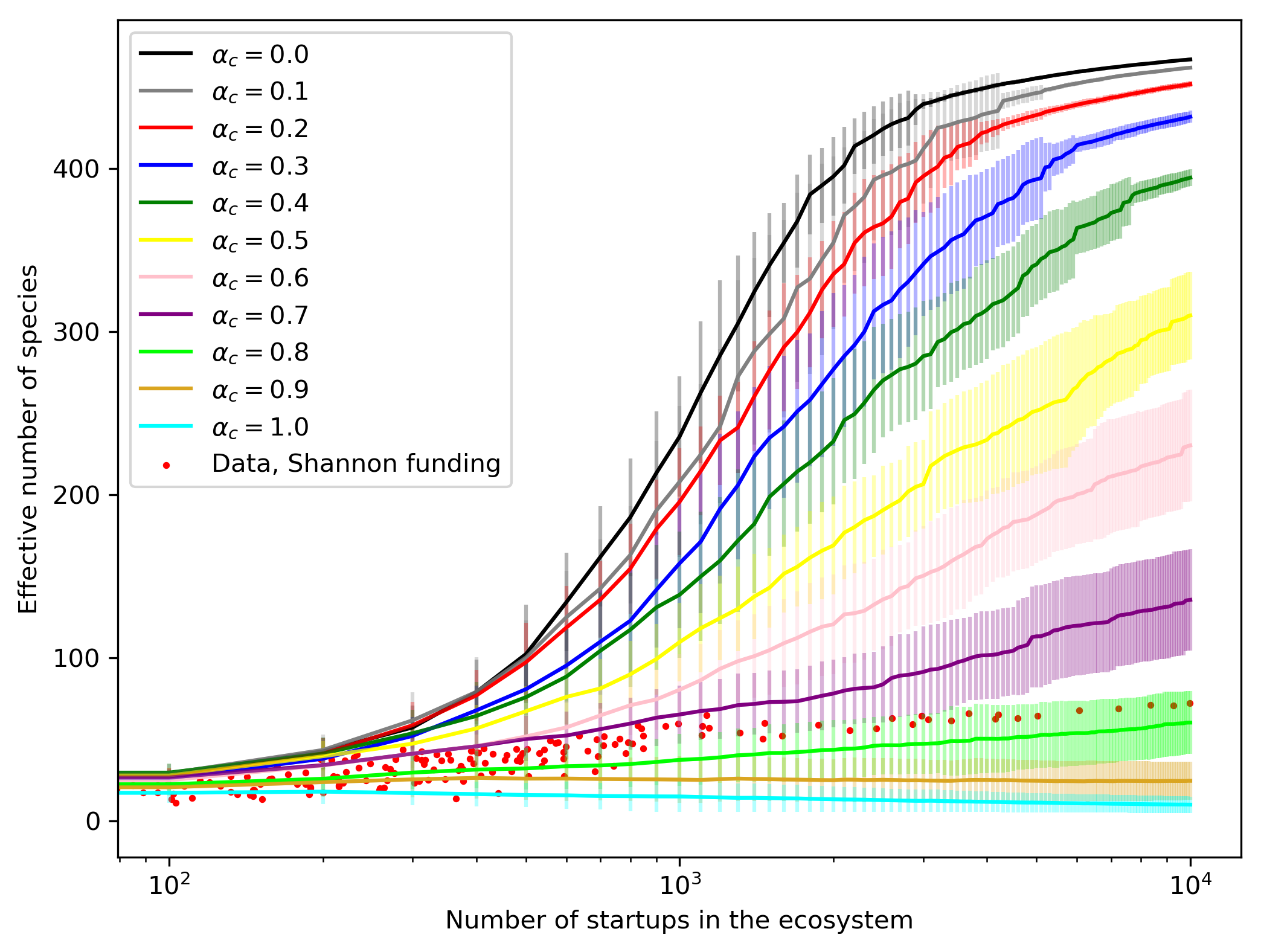}
    \caption{Simulation results for preferential attachment on number of startups (left) and on category funding (right). Diversity from the dataset was computed using Shannon funding.}
    \label{fig:PA_data-funding}
\end{figure}

\begin{figure}[!ht]
    \centering
    \includegraphics[scale = .4]{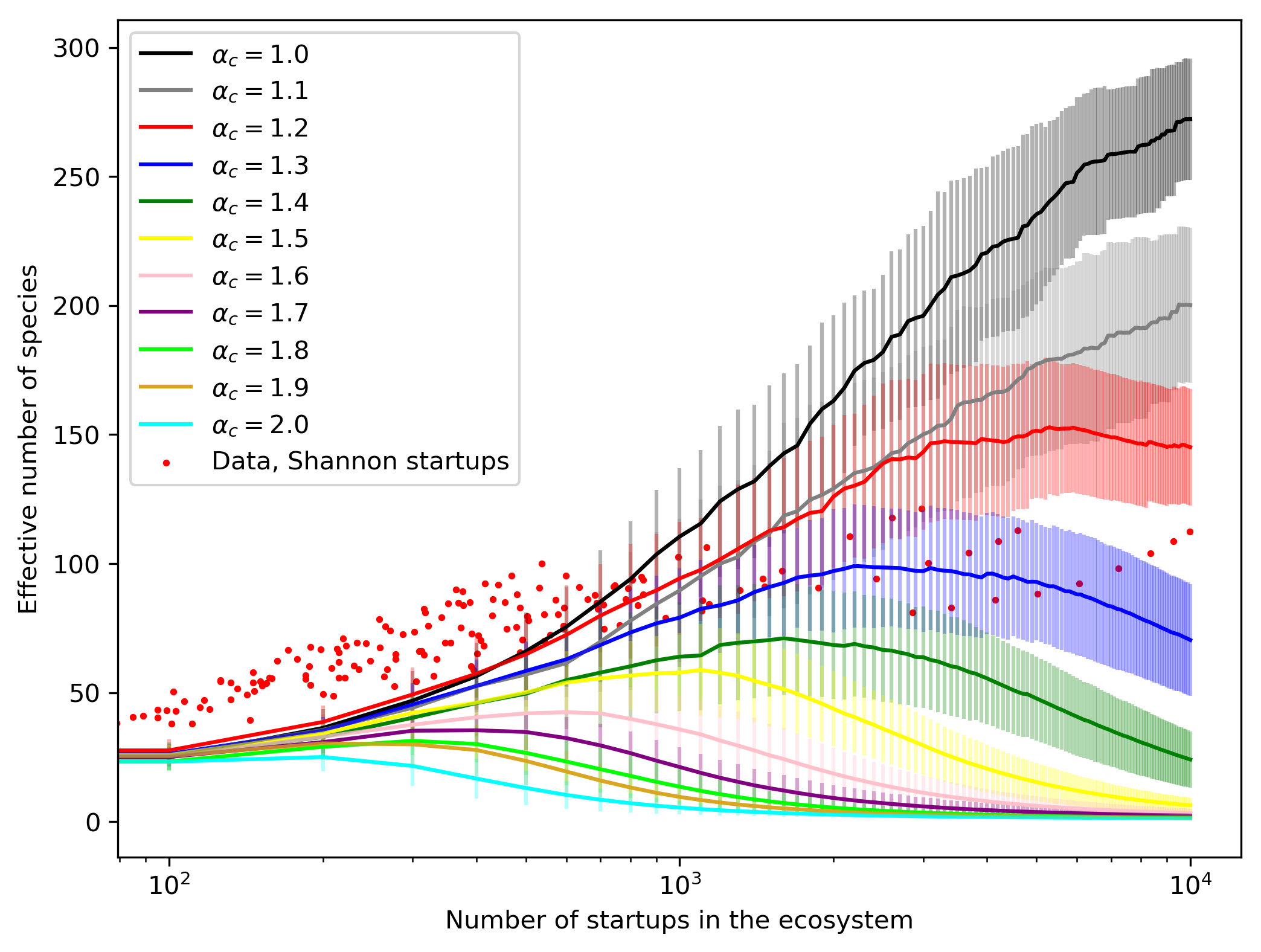}
    \includegraphics[scale = .4]{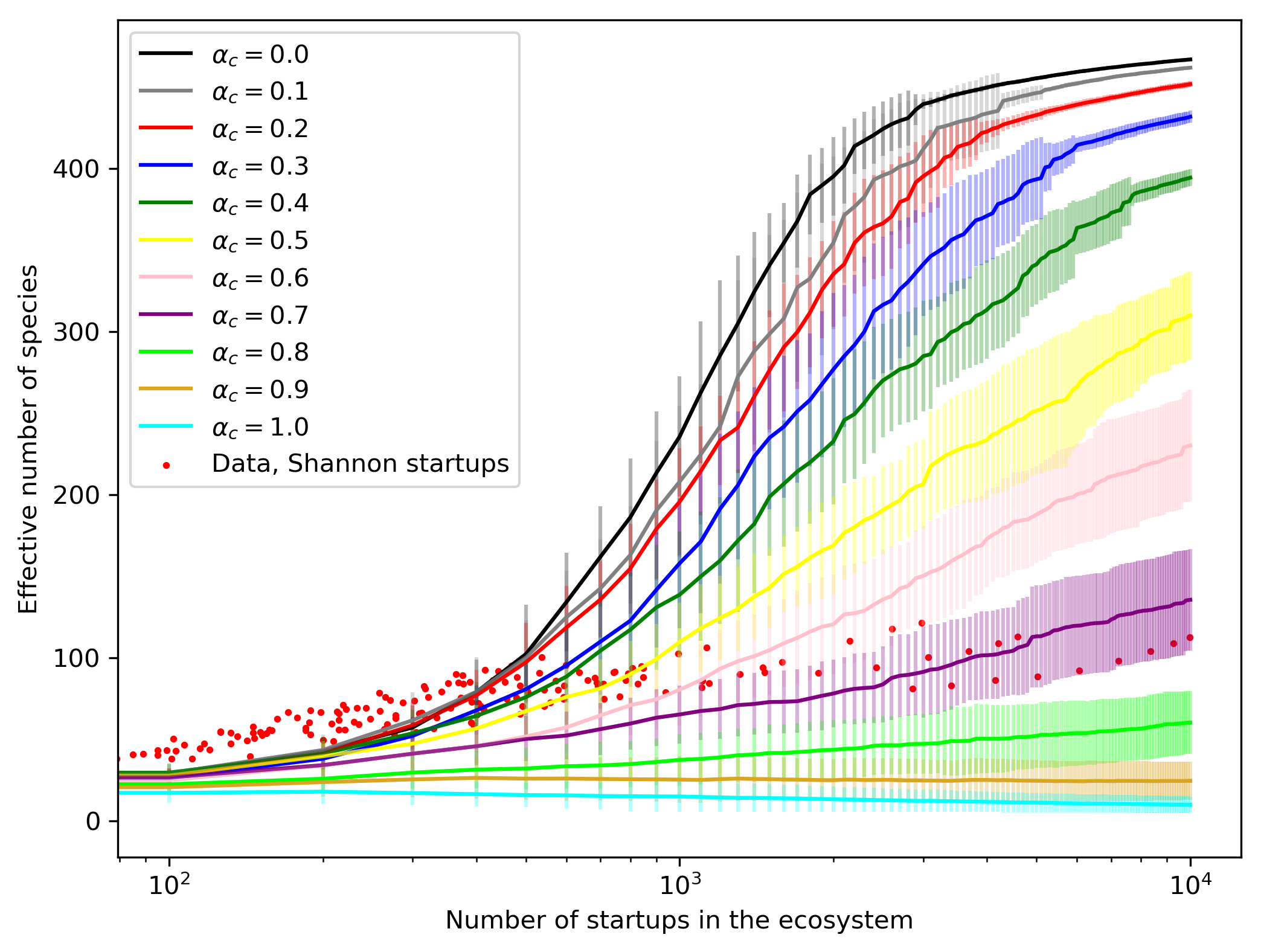}
    \caption{Simulation results for preferential attachment on number of startups (left) and on category funding (right). Diversity from the dataset was computed using Shannon startups.}
    \label{fig:PA_data-startups}
\end{figure}

Simulation results of the two variants of the model can be found in figs.~\ref{fig:PA_data-funding} and~\ref{fig:PA_data-startups}, with the diversity values from the simulation results (color lines) computed based on the entropy calculated using eq.~\ref{eq:entropy2} with the number of startups in each category (Shannon startups). We compared these values to diversity results from our dataset computed from the funding amounts (fig.~\ref{fig:PA_data-funding}) and the number of startups (fig.~\ref{fig:PA_data-startups}). These figures show that preferential attachment on the number of startups (left plots and eq.~\ref{eq:PA_startups}) seems to explain the diversification of the ecosystem up to a certain point, but that diversity is not stable as the ecosystem continues to grow \textit{i.e.} all new startups end up concentrating in a small number of categories and the effective number of species collapses. 

\begin{figure}[!ht]
    \centering
    \includegraphics[scale = .5]{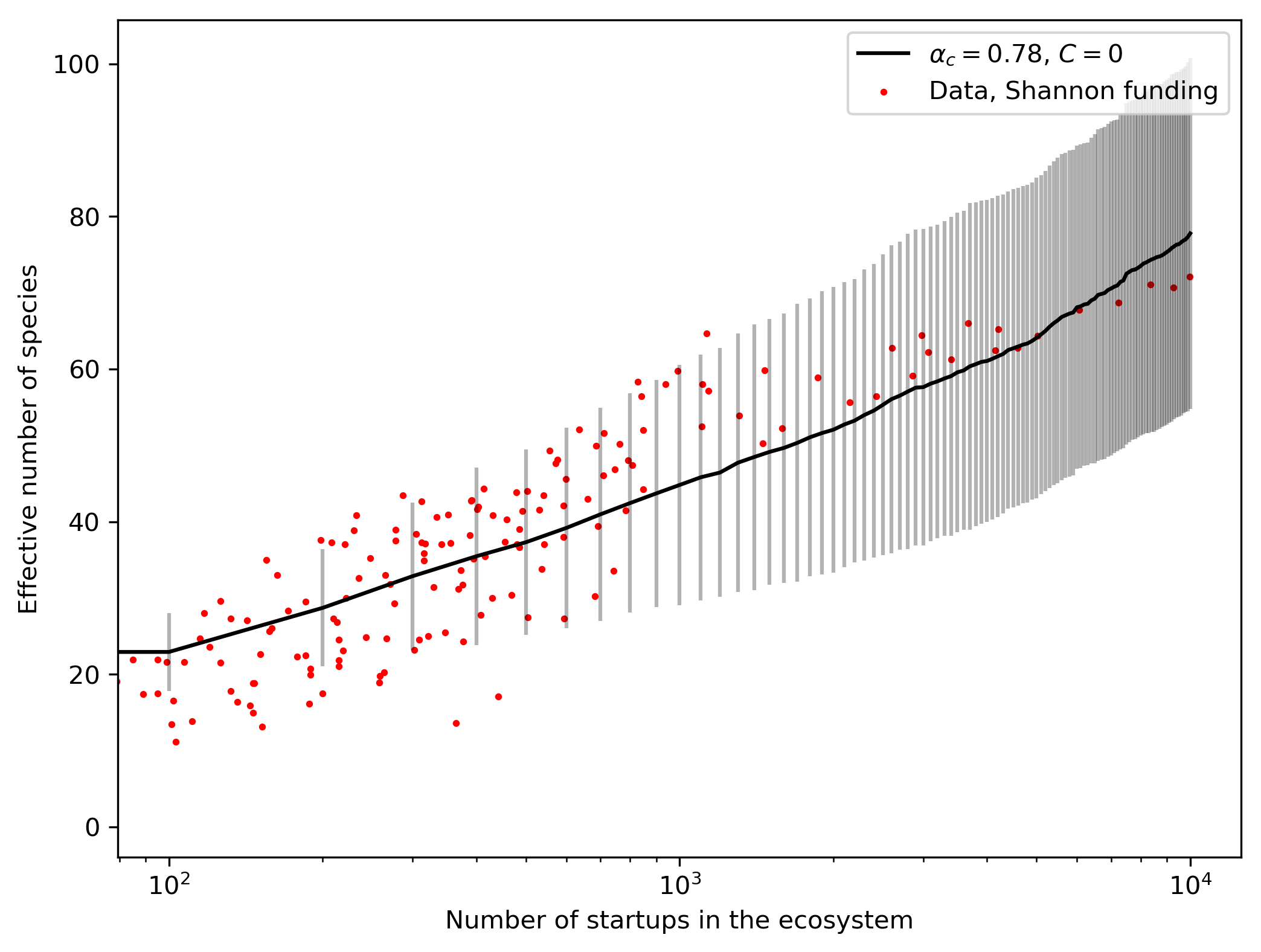}
    \caption{Effective number of species as a function of the number of startups in the ecosystem for preferential attachment on total funding with free parameter $\alpha_{c} = 0.78$. The black line corresponds to the simulation values averaged over 100 runs, the red points correspond to the data from our dataset. Error bars corresponding to one standard deviation are shown for the simulation.}
    \label{fig:sim_078}
\end{figure}

Preferential attachment on the category funding (right plots and eq.~\ref{eq:PA_funding}) on the other hand, seems to better match the data computed from Shannon startups and Shannon funding, as the ecosystem diversity steadily increases over time using this model. Preferential attachment on the funding amounts is thus a better mechanism than preferential attachment on the number of startups in order to explain the diversification of an ecosystem throughout its growth when comparing our results to the data. In the case of Shannon funding, the data and simulation results seem to match particularly well for free parameter $\alpha_{c}$ values around $0.8$ (see fig.~\ref{fig:sim_078}).

To check the robustness of these results, a numerical simulation variant of these models was tested where a new startup was placed in a random category with probability $p$ and with probability $1 - p$ was placed in a category following the preferential attachment law described in sec.~\ref{subsec:sim_procedure}. No qualitative differences were found between simulations with $0.1 < p < 0.25$ and $p = 0$ (the $p = 0$ case corresponds to a standard preferential attachment model as shown in figs.~\ref{fig:PA_data-funding}~and~\ref{fig:PA_data-startups}).
%\todo[inline, author=CG]{Est-ce qu'il y aurait moyen de mesurer la concordance des courbes?}

Fig.~\ref{fig:sim_078} shows that good concordance between the data points with Shannon funding (red dots) and simulation results (black line) is obtained for preferential attachment on funding with $\alpha_{c} = 0.78$.

Models of mixed preferential attachment taking into account both number of startups and total funding at the same time were tested following eq.~\ref{eq:hybrid_linear} : 

\begin{align}
    \label{eq:hybrid_linear}
    p_i = \frac{\theta (1 + n_{i})^{\alpha_{st}} + (1 - \theta)(\ln (x_{min}) + \ln  (f_{i}))^{\alpha_{funds}}}{\sum_{j = 1}^{N}[\theta (1 + n_{j})^{\alpha_{st}} + (1 - \theta)(\ln (x_{min}) + \ln (f_{j}))^{\alpha_{funds}}]}
\end{align}

with $\theta$ controlling the importance of funding amounts vs. number of startups and $\alpha_{st}$ and $\alpha_{funds}$ free parameters of the model. Simulations for a range of values of $\alpha_{st}$, $\alpha_{funds}$ and of $\theta \in [0, 1]$ did not provide a better match to the data than preferential attachment simply on the total category funding (fig.~\ref{fig:PA_data-funding}).

Finally, a simple mixed model of firm creation and growth was also confronted to the data. At each iteration of the simulation, a new startup is created with probability $\gamma$ in category $i$ following eq.~\ref{eq:PA_funding} and is allocated seed funding. With probability $1 - \gamma$, a random existing startup was funded with an amount depending on its last simulated funding round and moved on to the next stage of the "alphabet round" system (\textit{i.e.} a company that last received seed funding received series A funding, a company that last received series A funding received series B and so on). We set $\gamma = 0.5$ based on our data on the Silicon Valley ecosystem (seed rounds or "new entrants" represent approximately half of all venture funding rounds). The distribution of types of funding rounds at each stage with these parameters was found to be similar to that of our data. Simulation results from this mixed model of firm creation and growth did not give better results than the ones shown in fig.~\ref{fig:PA_data-funding}; the main driver between the diversification of an ecosystem then simply seems to be the allocation of fundings regardless of other ecosystem-dependant factors. The tendency of entrepreneurs to explore new industries or instead follow existing trends thus seems heavily linked to individual decisions which are particularly influenced by how financially successful the existing companies in the various categories have been.

\section{Conclusion}
In this paper, we presented a novel approach with respect to studying the emergence of startup ecosystems. Using public datasets, we first presented a novel, automated and interactive data visualization tool that facilitates the study of startup populations from an ecosystem point of view, and that also sheds light on the particularities of different ecosystems. Relatedly, diversity metrics such as the Shannon-Wiener index and the Simpson-Herfindahl index were then introduced, fostering the analogy with ecological sciences. We further tried to understand how observed diversity could emerge both by attempting to relate its disparity between ecosystems to macroeconomic indices and through numerical simulation. Our results suggest that the increase in diversity during the growth of a startup ecosystem can be explained through the sequential allocation of funding to startups in given sectors, thanks to a simple preferential attachment model, rather than by macro-economic indicators with the exception of economic development: i.e., startup ecosystem diversity appears as the outcome of emerging and aggregated behaviours rather than linked to ecosystem-specific characteristics or decisions. Needless to say, this analysis of ecosystem diversity remains preliminary and deserves further analysis, not only on a larger sample of ecosystems but also with a focus on events: for instance, linking "diversification" events, i.e. sectors getting a rather sudden and large influx of new startup creations, to specific "breakthroughs" -- either technological, as was recently the case with deep learning, or business-oriented, as has been seen with respect to \textit{Food Delivery} -- could typically give valuable insights into startup ecosystem diversity and diversification.

% coarse-grained as in the end, only two numbers $\{N(t), D(t)\}$ represent an ecosystem. It does not take into account the populated categories in terms of technological innovation and considers that breaking into the relatively simple (in terms of R\&D) \textit{Food Delivery} industry is not any easier or harder than breaking into a more technologically complex industry such as \textit{Energy}.

%Several fields of investigation can be imagined to complete this approach.
%More prospectively, a predictive algorithm could be developed to predict the probability of startups raising funds. 
%An approach involving job creation might allow public funding to be invested in most socially strategic startups.Further research is now needed to understand how those differences appeared and what it means for the startups and the cities hosting them.
%Finally, a dynamic approach to the problem could be considered in order to characterize the different states of development of ecosystems and how innovation policies influence them.

\newpage
\section*{Appendix}

\nolinenumbers

\end{document}